\documentclass[11pt]{article}
\usepackage{fullpage}
\usepackage{cite}
\usepackage{graphicx}
\usepackage{amsmath}
\usepackage{amssymb}
\title{}
\date{}

\begin{document}
\bibliographystyle{utphys}
\newcommand{\msbar}{\ensuremath{\overline{\text{MS}}}}
\newcommand{\DIS}{\ensuremath{\text{DIS}}}
\newcommand{\abar}{\ensuremath{\bar{\alpha}_S}}
\newcommand{\bb}{\ensuremath{\bar{\beta}_0}}
\newcommand{\rc}{\ensuremath{r_{\text{cut}}}}
\newcommand{\Nd}{\ensuremath{N_{\text{d.o.f.}}}}
\setlength{\parindent}{0pt}

\titlepage
\begin{flushright}
Edinburgh 2015/13 \\
\end{flushright}

\vspace*{0.5cm}

\begin{center}
{\Large \bf The classical double copy for Taub-NUT spacetime}
\end{center}

\vspace*{1cm}

\begin{center} 
\textsc{Andr\'{e}s Luna$^a$\footnote{a.luna-godoy.1@research.gla.ac.uk}, 
Ricardo Monteiro$^b$\footnote{monteiro@maths.ox.ac.uk}, 
Donal O'Connell$^{c,d}$\footnote{donal@staffmail.ed.ac.uk} and 
Chris D. White$^a$\footnote{Christopher.White@glasgow.ac.uk} } \\

\vspace*{0.5cm} $^a$ School of Physics and Astronomy, University of Glasgow,\\ Glasgow G12 8QQ, Scotland, UK\\

\vspace*{0.5cm} $^b$ Mathematical Institute, University of Oxford, \\Oxford OX2 6GG, England, UK\\

\vspace*{0.5cm} $^c$ Higgs Centre for Theoretical Physics, \\School of Physics and Astronomy, The University of Edinburgh,\\
Edinburgh EH9 3JZ, Scotland, UK\\

\vspace*{0.5cm} $^d$ Kavli Institute for Theoretical Physics,
University of California,\\
Santa Barbara, CA 93106-4030 USA\\

\end{center}

\vspace*{0.5cm} 

\begin{abstract}
The double copy is a much-studied relationship between gauge theory
and gravity amplitudes. Recently, this was generalised to an infinite
family of classical solutions to Einstein's equations, namely
stationary Kerr-Schild geometries. In this paper, we extend this to
the Taub-NUT solution in gravity, which has a double Kerr-Schild
form. The single copy of this solution is a dyon, whose electric and
magnetic charges are related to the mass and NUT charge in the gravity
theory. Finally, we find hints that the classical double copy extends
to curved background geometries.
\end{abstract}

\vspace*{0.5cm}

\section{Introduction}
\label{sec:intro}
The study of scattering amplitudes in both gauge and gravity theories
continues to be an active area of research, including the
relationships between both types of theory. A relatively recent
discovery is the {\it double copy} property linking gauge and gravity
amplitudes, with or without
supersymmetry~\cite{Bern:2008qj,Bern:2010yg,Bern:2010ue}. This itself
relies on a certain relationship ({\it BCJ duality}) being made
manifest in the gauge theory, and is known to be true at
tree-level~\cite{BjerrumBohr:2009rd,Stieberger:2009hq,Bern:2010yg,BjerrumBohr:2010zs,Feng:2010my,Tye:2010dd,Mafra:2011kj,BjerrumBohr:2012mg},
where it is equivalent to the KLT relations
of~\cite{Kawai:1985xq}. Follow-up work has examined loop-level
amplitudes~\cite{Bern:2010ue,Bern:1998ug,Green:1982sw,Bern:1997nh,Oxburgh:2012zr,Carrasco:2011mn,Carrasco:2012ca,Bargheer:2012gv,Mafra:2012kh,Boels:2013bi,Bjerrum-Bohr:2013iza,Bern:2013yya,Bern:2013qca,Nohle:2013bfa,Bern:2013uka,Naculich:2013xa,Du:2014uua,Mafra:2014gja,Bern:2014sna,Mafra:2015mja},
form-factors~\cite{Boels:2012ew}, or alternative
theories~\cite{Johansson:2014zca}. All-order tests of the double copy
are possible in certain kinematic
limits~\cite{Oxburgh:2012zr,Saotome:2012vy,Vera:2012ds,Johansson:2013nsa,Johansson:2013aca},
but a full proof at loop level -- which relies on the existence of
BCJ-dual amplitudes -- has yet to be obtained (see
refs.~\cite{Monteiro:2013rya,Tolotti:2013caa,Fu:2013qna,Du:2013sha,Fu:2012uy,Naculich:2014naa,Naculich:2014rta,Chiodaroli:2014xia,Carrasco:2013ypa,Litsey:2013jfa,Nagy:2014jza,Weinzierl:2014ava,Ho:2015bia,Anastasiou:2015vba,Johansson:2015oia,Barnett:2014era}
for related studies). It is not known, for example, if the copy is a
genuinely non-perturbative property of both theories. One reason for
this is that the perturbative construction of the double copy (which
relies on replacing four-gluon vertices with pairs of three-gluon
vertices) obscures a direct understanding at the level of the
Lagrangian. Partial exceptions to this are the perturbatively
constructed effective Lagrangian of refs.~\cite{Bern:2010yg} (see also
refs.~\cite{Tolotti:2013caa,Vaman:2014iwa}), and the self-dual sector
of refs.~\cite{Monteiro:2011pc,Boels:2013bi} where, in the latter, the
Yang-Mills action can be made manifestly cubic~\cite{Parkes:1992rz},
and a full interpretation of BCJ duality obtained. See
Ref.~\cite{Carrasco:2015iwa} for a recent review. \\

The above discussion motivates an alternative way of examining the
double copy, namely to look directly at solutions of the classical
equations of motion in gauge and gravity theories, and to match these
up according to a double copy prescription. If this matching can be
argued to be the same as the BCJ double copy for amplitudes in a
meaningful way, one gains a deeper understanding of the double copy,
including its potentially nonperturbative role. A pioneering study in
this regard was recently undertaken by some of the present authors in
ref.~\cite{Monteiro:2014cda}, which found an infinite class of
classical solutions that double copy between Yang-Mills theory and
gravity. On the gravity side, these correspond to stationary
Kerr-Schild solutions, where the latter property leads to
linearisation of the Einstein equations, such that the solution for
the graviton field terminates at first-order in perturbation
theory. The single copies of these solutions solve linearised
Yang-Mills equations, and a number of examples were given in
ref.~\cite{Monteiro:2014cda}. That the single copy procedure is indeed
related to the BCJ double copy for amplitudes relies on performing the
zeroth copy from the gauge theory to a biadjoint scalar theory. The
latter has arisen in a number of recent
studies~\cite{Bern:1999bx,BjerrumBohr:2012mg,Cachazo:2013iea,Anastasiou:2014qba},
and its relevance in the present context is that the scalar field from
the zeroth copy can be used to fix the double copy prescription
between gauge theory and gravity. It was also shown in
ref.~\cite{Monteiro:2014cda} that the self-dual sectors of gauge
theory and gravity have a Kerr-Schild-like description.\\

A number of puzzles remain regarding the results of
ref.~\cite{Monteiro:2014cda}, not least the lack of a full
understanding of the role that Kerr-Schild coordinates play. The aim
in this paper is to generalise the results of
ref.~\cite{Monteiro:2014cda}, and to provide further evidence
supporting the classical double copy. To this end, we consider the
Taub-NUT solution~\cite{Taub,NUT} in General Relativity, which is
known not to have a simple Kerr-Schild form. Nevertheless, we will
obtain a single copy of this solution, which has a clear
interpretation (in four spacetime dimensions) as a gauge-theory
dyon. As for the simple Kerr-Schild solutions of
ref.~\cite{Monteiro:2014cda}, one may also take a zeroth copy to a
biadjoint scalar theory, which can be used to fix the nature of the
double copy, and to relate this to the double copy for amplitudes. The
Schwarszchild black hole emerges as a special case.\\

Another generalisation is to consider Kerr-Schild geometries that have
a non-trivial (i.e. non-Minkowski) background metric. We examine the
Taub-NUT solution on (Anti)-de Sitter space, and find that the single
copy also works in this case, namely that one may construct a gauge
theory object that satisfies the curved space Maxwell equations. We
find an extra term in the biadjoint field equation, which is
consistent with the scalar being conformally coupled to the gravity
background in four spacetime dimensions.\\

The structure of our paper is as follows. In section~\ref{sec:KSreview},
we review the results of ref.~\cite{Monteiro:2014cda}. In
section~\ref{sec:taubnut} we discuss the Taub-NUT solution and its
gauge theory counterpart, and in section~\ref{sec:curved} we look at
the double copy in (Anti)-de Sitter space. We discuss our results and
conclude in section~\ref{sec:conclude}.

\section{The Kerr-Schild double copy}
\label{sec:KSreview}

A special family of solutions to Einstein's gravitational field
equations comprises {\it Kerr-Schild} metrics (see
e.g.~\cite{Stephani:2003tm} for a detailed review)
\begin{align}
g_{\mu\nu}&=\bar{g}_{\mu\nu}+\kappa h_{\mu\nu}\notag\\
&=\bar{g}_{\mu\nu}+\kappa\,\phi\, k_\mu\, k_\nu,
\label{KSdef}
\end{align}
where $\kappa=\sqrt{16\pi G_N}$, $G_N$ is Newton's constant, and
$\bar{g}_{\mu\nu}$ is a background metric. Here $\phi$ is a scalar
field, and $k_\mu$ is both null and geodetic with respect to the
background. That is,
\begin{equation}
\bar{g}_{\mu\nu}\, k^{\mu}\, k^\nu=0,\quad (k\cdot D)\, k_\mu=0,
\label{kconditions}
\end{equation}
where $D^\mu$ is the covariant derivative in the metric
$\bar{g}_{\mu\nu}$. The Kerr-Schild form is special in that the
``graviton" explicitly decomposes into an outer product of the vector
$k_\mu$ with itself. Furthermore, such solutions have the remarkable
property that they linearise the Einstein equations. More
specifically, the components of the Ricci tensor
are~\footnote{N.B. the Ricci tensor is only linearised for the index
  placement as chosen in eq.~(\ref{Ricci}).}
\begin{equation}
R^\mu_\nu=\bar{R}^\mu_\nu-\kappa \left[h^{\mu}_\rho\bar{R}^\rho_\nu
+\frac{1}{2} D_\rho\left(D_\nu h^{\mu\rho}+D^\mu h^\rho_\nu
-D^\rho h^\mu_\nu\right)\right],
\label{Ricci}
\end{equation}
where $\bar{R}_{\mu\nu}$ is the Ricci tensor associated with
$\bar{g}_{\mu\nu}$. Reference~\cite{Monteiro:2014cda} concentrated on
a Minkowski background, $\bar{g}_{\mu\nu}=\eta_{\mu\nu}$, and also
stationary solutions ($\partial_0\phi=\partial_0 k_\mu=0$). It was then
shown that the gauge field
\begin{equation}
A^a_\mu=c^a\phi\,k_\mu,
\label{singlecopy}
\end{equation}
for constant $c^a$, solves the Yang-Mills equations, which due to the trivial colour dependence take a
linearised form,
\begin{equation}
\partial^\mu F^a_{\mu\nu}=0.
\label{YMlin}
\end{equation}
This gauge field of eq.~(\ref{singlecopy}) is then taken to be the
single copy of the graviton of eq.~(\ref{KSdef}), obtained by removing
one of the Lorentz vectors $k_\mu$ from the graviton solution, and
replacing coupling constant and charge factors. Carrying this one step
further, one may remove the additional Lorentz factor, and define the
biadjoint scalar field
\begin{equation}
\Phi^{aa'}=c^a \tilde{c}^{a'}\phi,
\label{biadjointlin}
\end{equation}
for constant colour charge vectors $c^a$ and $\tilde{c}^{a'}$, where $a$ and $a'$ are associated to the Lie algebras of two distinct groups $G$ and $G'$. This
solves the equation of motion for the biadjoint scalar
theory (which linearises),
\begin{equation}
\partial^2\Phi^{aa'}-yf^{abc}\tilde{f}^{a'b'c'}\Phi^{bb'}\Phi^{cc'}=0,
\label{biadjoint}
\end{equation}
and is then identified with the zeroth copy of the gauge theory
solution. The zeroth copy is one way of fixing the single copy
procedure of eq.~(\ref{singlecopy}), which would otherwise be
ambiguous. A priori, one can choose to absorb an overall scalar
function into the Kerr-Schild vector $k_\mu$ before taking the single
copy. However, there is a unique choice that satisfies the biadjoint
equation upon taking the zeroth copy. Furthermore, this has a physical
interpretation as a scalar propagator integrated over the source
charge~\cite{Monteiro:2014cda}. This is the same procedure as the BCJ
double copy for amplitudes, in which denominator factors (scalar
propagators) are left untouched when the double copy is performed, but
numerators are not. \\

\section{The Taub-NUT solution}
\label{sec:taubnut}

The Taub-NUT metric, first derived in refs.~\cite{Taub,NUT}, is a
stationary, axisymmetric vacuum solution that is not asymptotically
flat. It can be sourced by a pointlike object at the origin, which
carries both electric charge and {\it NUT charge}. The latter is
associated with the lack of spherical symmetry and asymptotic
flatness, and is known to correspond to a magnetic monopole-like
behaviour of the gravitational field at spatial infinity (see
e.g.~\cite{Ortin:2004ms} for a review). We will see in this section
that the Taub-NUT solution provides an interesting example of a
Kerr-Schild double copy, which extends our previous results to metrics
not of the form of eq.~(\ref{KSdef}).\\

A general formulation of the Taub-NUT-Kerr-de Sitter metric has been
given by Plebanski~\cite{Plebañski1975196}, and has later been
shown to exhibit a {\it double Kerr-Schild form} in
ref.~\cite{Chong:2004hw}. That is, one may write the metric in the form
\begin{align}
g_{\mu\nu}&=\bar{g}_{\mu\nu}+\kappa h_{\mu\nu}\notag\\
&=\bar{g}_{\mu\nu}+\kappa\left(\phi\,k_\mu\,k_\nu+\psi\,l_\mu\,l_\nu
\right),
\label{doubleKS}
\end{align}
where $\bar{g}_{\mu\nu}$ is a de Sitter background metric, the vectors
$k_\mu$ and $l_\mu$ satisfy the conditions
\begin{equation}
k^2=l^2=k\cdot l=0,\quad (k\cdot D)k_\mu=0,\quad (l\cdot D)l_\mu=0,
\label{klconditions}
\end{equation}
and all contractions and covariant derivatives can be taken with
respect to either the background or the full metric. This form is a
clear generalisation of the standard Kerr-Schild form of
eq.~(\ref{KSdef}). However, it is not a straightforward extension: it
is no longer true in general that the Einstein equations are
linearised. One may show, in fact, that the mixed components of the
Ricci tensor are
\begin{equation}
R^\mu_\nu=\bar{R}^\mu_\nu+\kappa\left[-h^\mu_\rho\bar{R}^\rho_\nu
+\frac{1}{2}D_\rho\left(D_\nu h^{\mu\rho}+D^\mu h_\nu^\rho-D^\rho 
h_\nu^\mu\right)\right]+R^\mu_{\nu, {\rm non-lin.}},
\label{RiccidoubleKS}
\end{equation}
where the non-linear term is
\begin{align}
R^\mu_{\nu, {\rm non-lin.}}&=-\frac{\kappa^2}{2}\left[\frac{1}{2}
D^\mu h(k)^\rho_\delta D_\nu h(l)^\delta_\rho
+h(l)^{\mu\delta}D_\rho D_\nu h(k)^\rho_\delta\right.\notag\\
&\left.\phantom{\frac{1}{2}}+D_\rho\left(
h(l)^{\rho\delta}D_{\delta}h(k)^{\mu}_\nu+2h(l)^{\rho\delta}D_{(\nu}h(k)^{\mu)}_\delta-2h(l)^{\mu\delta}
D^{[\rho}h(k)_{\delta]\nu}\right)\right]+(k\leftrightarrow l),
\label{Rnonlin}
\end{align}
and we have defined the shorthand notation
\begin{equation}
h(k)_{\mu\nu}=\phi k_\mu k_\nu,\quad h(l)_{\mu\nu}=\psi l_\mu l_\nu.
\label{hkhl}
\end{equation}
Remarkably, in Plebanski coordinates, as argued in
ref.~\cite{Chong:2004hw}, the full metric can be cast in a form such
that eq.~(\ref{Rnonlin}) vanishes, and the Ricci tensor indeed
linearises. The explicit form of the background line element is
\begin{align}
d\bar{s}^2=-\frac{1}{q^2-p^2}\left[\bar{\Delta}_p(d\tilde{\tau}
+q^2d\tilde{\sigma})^2+\bar{\Delta}_q(d\tilde{\tau}+p^2d\tilde{\sigma})^2
\right]-2(d\tilde{\tau}+q^2d\tilde{\sigma})dp-2(d\tilde{\tau}+p^2d\sigma)dq,
\label{gibbonspope}
\end{align}
where
\begin{equation}
\bar{\Delta}_p=\gamma-\epsilon p^2+\lambda p^4,\quad 
\bar{\Delta}_q=-\gamma+\epsilon q^2-\lambda q^4.
\label{Deltapq}
\end{equation}
Here $\epsilon$ is a constant, and $\gamma$ is related to the angular
momentum. Equation~(\ref{gibbonspope}) is a solution to the Einstein
equation with non-zero cosmological constant $\lambda$. The
Kerr-Schild vectors are given in the
$(\tilde{\tau},\tilde{\sigma},p,q)$ coordinate system (which has (2,2)
signature) by
\begin{equation}
k_\mu=(1,q^2,0,0),\quad l_\mu=(1,p^2,0,0),
\label{kldef}
\end{equation}
and the accompanying scalar functions by
\begin{equation}
\phi=\frac{2Np}{q^2-p^2},\quad \psi=\frac{2mq}{q^2-p^2}.
\label{phipsi}
\end{equation}
The parameter $m$ represents the mass of the solution, and $N$ the NUT
charge. \\

Let us now obtain and interpret the single copy of this solution,
where we will focus for brevity on the case of vanishing angular momentum
($\gamma=0$), which leads to a pointlike source. First, we note that
the natural generalisation of the single copy procedure of
eq.~(\ref{singlecopy}) in the double Kerr-Schild case is to construct
a gauge field
\begin{equation}
A^a_\mu=c^a\left(\phi k_\mu+\psi l_\mu\right).
\label{singlecopydouble}
\end{equation}
That is, the double copy of this solution proceeds term-by-term,
analagously to how the BCJ double copy for amplitudes is applied
separately to terms involving different scalar propagators. We have
verified that the gauge field of eq.~(\ref{singlecopydouble})
satisfies the Yang-Mills equations (which linearise),
\begin{equation}
D^\mu F^a_{\mu\nu}=0,\quad F^a_{\mu\nu}=D_\mu A_{\nu}^a
-D_\nu A_\mu^a.
\label{Maxwell}
\end{equation}
It is interesting already at this point that
the Yang-Mills equations are satisfied even for a non-Minkowski
background. We return to this in the following section.\\

For now, let us interpret the case $\lambda=0$, for which
$\bar{g}_{\mu\nu}=\eta_{\mu\nu}$. In taking the single copy, we will
make the replacements
\begin{equation}
\frac{m\kappa}{2}\rightarrow (c_aT^a)g_s,\quad \frac{N\kappa}{2}\rightarrow
(c_aT^a)\tilde{g}_s,
\label{mreplace}
\end{equation}
to be explained shortly. Having taken the single copy in Plebanksi
coordinates (where the Kerr-Schild form is manifest), we are free to
transform to any other choice of coordinates. We will do this in two
stages. First, following refs.~\cite{Plebañski1975196,Chong:2004hw},
one may transform to spheroidal coordinates according to
\begin{equation}
\tau=t+a\varphi, \quad \sigma=\frac{\varphi}{a},\quad
q=r, \quad p=a\cos\theta,
\label{trans}
\end{equation}
where the coordinates $\tau$ and $\sigma$ are related to the
counterparts used throughout this paper by
\begin{equation}
d\tilde{\tau}=d\tau+\frac{p^2dp}{\Delta_p}-\frac{q^2dq}{\Delta_q},\quad
d\tilde{\sigma}=d\sigma-\frac{dp}{\Delta_p}+\frac{dq}{\Delta_q}.
\label{tildetrans}
\end{equation}
Next, one may take the parameter $a^2\equiv\gamma$ (which is related
to the angular momentum) to zero, so that the spheroidal radius
becomes a spherical one. This coordinate transformation is subtle, in
that the vector $l^\mu$ becomes singular as $a\rightarrow 0$. The
prefactor $\psi$ entering the gauge field, however, is ${\cal O}(a)$,
such that gauge field $A_\mu^a$ itself is well-defined. In the
spherical polar coordinate system $(t,r,\theta,\phi)$, the field
strength tensor then becomes
\begin{equation}
F= \frac{1}{2}F_{\mu\nu}\,dx^\mu \wedge dx^\nu=-\frac{c_aT^a}{8\pi} \, \left( \frac{g_s}{r^2} \,dt \wedge dr+ \tilde{g}_s \,\sin \theta\, d\theta \wedge d\phi \right) ,
\label{Fmunu2}
\end{equation}
where we have separated out the contributions from the constants $g_s$
and $\tilde{g}_s$. The first term on the right-hand side of
eq.~(\ref{Fmunu2}) gives a pure electric field, corresponding to a
Coulomb solution. Thus, the mass in the Taub-NUT metric single copies
to a static colour charge, exactly as in the Schwarzschild case of
ref.~\cite{Monteiro:2014cda}. This must in fact be the case, given
that the Taub-NUT metric becomes the Schwarzschild metric as
$N\rightarrow 0$. This explains our choice of factors in
eq.~(\ref{mreplace}). \\

The NUT charge contribution to the field strength tensor is a pure
magnetic field, and we can interpret this in more detail by expressing
eq.~(\ref{Fmunu2}) as
\begin{equation}
F=-\frac{c_aT^a}{8\pi} \, \left( \frac{g_s}{r^2} \,dt \wedge dr+ \star\, \frac{\tilde{g}_s}{r^2} \,dt \wedge dr \right) .
\label{Ftildel}
\end{equation}
Here, $\star$ denotes the Hodge dual of a 2-form, say $\omega_{\mu\nu}$,
\begin{equation}
\star\, \omega_{\mu\nu}: =\frac{1}{2}\epsilon_{\mu\nu\alpha\beta}\, \omega^{\alpha\beta},
\qquad \star^2=-1.
\label{Ftilde}
\end{equation}
Thus, the dual tensor for the NUT-charge term contains a pure electric
field corresponding to a point charge of strength $\tilde{g}_s$. It
follows that the magnetic field in the original field strength tensor
corresponds to a magnetic monopole, where the NUT charge in the
gravity theory single copies to the monopole charge in the gauge
theory. This is perhaps to be expected, given that the NUT charge in
the Taub-NUT metric is known to be associated with monopole-like
behaviour~\cite{Ortin:2004ms}, an analogy which has now been turned
into an exact statement under the classical double copy. We have then
chosen the constant $\tilde{g}$ in eq.~(\ref{mreplace}) to obey the
same normalisation as $g_s$ in the (non-dual) field strength tensor.\\

Note that the transformation from the Plebanski coordinate system to
the spherical coordinate system involves a change of signature (from
(2,2) to (1,3)), and thus a Wick rotation. In the Plebanski system
itself, the two charges $m$ and $l$ appear on an equal footing, as is
clear from eqs.~(\ref{gibbonspope}--\ref{phipsi}). In other words, in
this signature one cannot tell the difference in the gauge theory
between an electric and a (dual) magnetic charge. For the (anti-)
self-dual case, the gauge and gravity solutions can be interpreted as
instantons (see also~\cite{Ortin:2004ms}).\\

As is well known, consistency of the monopole gauge field leads to the
quantisation condition (in the present notation)
\begin{equation}
g\tilde{g}=\frac{n}{2},\quad n\in\mathbb{Z},
\label{quantcondition}
\end{equation}
relating the electric and magnetic charges. This has an analogue in
the gravity theory, as discussed in ref.~\cite{Misner,Dowker}. There,
recovery of spherical symmetry demands a periodic time
coordinate. This corresponds to quantisation of the energy of the
dyon, or its mass in the non-relativistic approximation. There is then
a quantisation condition relating the mass and NUT charge, which is
the equivalent of eq.~(\ref{quantcondition}) from a double copy
perspective.\\

As in the standard Kerr-Schild case, we may take the zeroth copy, which
produces a biadjoint scalar field 
\begin{equation}
\Phi^{aa'}=c^a\tilde{c}^{a'}\left(\phi+\psi\right).
\label{biadjointtaubnut}
\end{equation}
Similarly to the results of ref.~\cite{Monteiro:2014cda}, this is a
solution of the linearised biadjoint eq.~(\ref{biadjoint}). In fact, both $\phi$ and $\psi$ satisfy that equation separately. They have the interpretation of a scalar propagator integrated
over the source charges, and are analagous to the scalar propagators
that are not modified when double-copying scattering amplitudes. As
has already been mentioned above, another property that links the
generalised Kerr-Schild double copy to the corresponding story for
amplitudes is that each Kerr-Schild term (involving a different scalar
propagator) is copied individually, with no mixing between these terms
on the gravity side.\\

The results of this section constitute an interesting generalisation
of the Kerr-Schild double copies of ref.~\cite{Monteiro:2014cda}, in
that a double Kerr-Schild form is used. As remarked in
ref.~\cite{Chong:2004hw}, it is highly non-trivial that the particular
double Kerr-Schild result for the Taub-NUT solution linearises the
Einstein equations. One may also examine analogues of the Taub-NUT
solution in higher dimensions. A family of higher-dimensional
generalisations of the Plebanski metric was obtained in
ref.~\cite{Chen:2006xh}, and subsequently shown to have a multiple
Kerr-Schild form~\cite{Chen:2007fs}, involving $n=\lfloor D/2 \rfloor$
linearly independent, mutually orthogonal null vector fields $k_i^\mu$
(here $\lfloor X \rfloor$ denotes the integer part of $X$):
\begin{equation}
g_{\mu\nu}=\bar{g}_{\mu\nu}+\sum_{i=1}^n \phi_ik_{i\mu}k_{i\nu}.
\label{higherdimTN}
\end{equation}
This form relies on a generalised set of Plebanski-like coordinates,
in $\left( \lfloor D/2\rfloor,\lfloor (D+1)/2\rfloor\right)$ signature, and
each function $\phi_i$ involves a parameter playing the role of a
generalised NUT charge. Assuming that these metrics indeed linearise
the Einstein equations~\footnote{The linearisation property is not
  explicitly stated in ref.~\cite{Chen:2007fs}. However, we have
  checked its validity up to $D=7$.}, one may construct a single copy
gauge field term by term, as for the double Kerr-Schild example in
$D=4$:
\begin{equation}
A_\mu=\sum_{i=1}^n\phi_i k_{i\mu}.
\label{higherdimTN2}
\end{equation} 
Each term taken by itself satisfies the Maxwell equations using the
argument of ref.~\cite{Monteiro:2014cda}, given that it constitutes a
time-independent single Kerr-Schild solution. That the complete
multiple Kerr-Schild form satisfies the Maxwell equations then follows
from linear independence of the generalised NUT charges. Note that, in
the signature of the Plebanski-like metric, the NUT charges appear on
an equal footing, as in the canonical Taub-NUT case. After analytic
continuation to the physical (1,$D-1$) signature, one parameter will
play the role of an electric charge, obtained as a single copy of a
mass parameter in the gravity theory.\\

As mentioned above, there have been previous observations that the
Taub-NUT solution is analagous to a gauge theory
dyon~\cite{Dowker,Ortin:2004ms}. Such statements, however, are
restricted to the weak gravity approximation, and are not embedded in
a formal double copy relationship between Yang-Mills theory and
gravity. Here, the use of the Kerr-Schild double copy makes the
dyon-Taub-NUT relationship perturbatively exact, and also ties it to
the double copy for scattering amplitudes. 

\section{Double copy in de Sitter space}
\label{sec:curved}

In the previous section, we saw that the Kerr-Schild double copy can
be extended to the case of a double Kerr-Schild solution, representing
a dyon. Another possible generalisation is to consider the background
metric $\bar{g}_{\mu\nu}$ to be non-Minkowski, and the Plebanski form
of the Taub-NUT-Kerr-de Sitter metric provides just such an example.\\

We have, in fact, already seen above that the gauge field of
eq.~(\ref{singlecopydouble}), obtained as a single copy of the
Plebanski metric, solves the {\it (Anti-) de Sitter space} Maxwell
equations of eq.~(\ref{Maxwell}). This already suggests that our
interpretation of the Kerr-Schild double copy can indeed be
generalised to curved backgrounds. In the Minkowski case, it was
important to take the zeroth copy to a biadjoint scalar theory. This
fixes the overall scalar function that is not squared when taking the
double copy, and also helps to tie the classical double copy to the
similar procedure for scattering amplitudes. One may then also examine
the zeroth copy in the (Anti-) de Sitter background. For the general
Taub-NUT-Kerr-de Sitter metric, one finds that the scalar field of
eq.~(\ref{biadjointtaubnut}) satisfies
\begin{equation}
D^2\Phi^{aa'}=-2\lambda \Phi^{aa'}.
\label{biadjointdS}
\end{equation}
This has an additional term on the RHS, and it is not immediately
clear how to interpret this. However, it is intriguing to note that this term
is in fact proportional to the Ricci curvature, such that
eq.~(\ref{biadjointdS}) is obtained from the Lagrangian:
\begin{equation}
{\cal L}=\frac{1}{2}(D^\mu\Phi^{aa'})(D_\mu\Phi^{aa'})-\frac{y}{6}f^{abc}
\tilde{f}^{a'b'c'}\Phi^{aa'}\Phi^{bb'}\Phi^{cc'}
-\frac{\cal R}{12}\Phi^{aa'}\Phi^{aa'},
\label{Lcurved}
\end{equation}
corresponding to a non-minimal coupling of the biadjoint scalar to the
gravity background. More than this, the coefficient of the extra term precisely
coincides with a conformally coupled scalar in four spacetime
dimensions. This perhaps can be explained from the fact that classical
Yang-Mills theory is conformally invariant in four spacetime
dimensions, and that the zeroth copy somehow preserves this invariance
in the free scalar theory.\footnote{In higher dimensions, Yang-Mills
  theory is not conformally invariant, and the relevant coefficient of
  the ${\cal R} \Phi^{aa'}\Phi^{aa'}$ term does not coincide with the
  conformal coupling.} Whether or not this is the correct
interpretation of this result, deserves further investigation.\\

It should be mentioned that it is possible to reinterpret the de
Sitter double copy as a multiple Kerr-Schild double copy around
Minkowski space. This is because the de Sitter metric itself can be
written in the Kerr-Schild form~\cite{Chen:2007fs}
\begin{equation}
g_{{\rm dS}, \mu\nu}=\eta_{\mu\nu}+\lambda\, r^2 n_\mu n_\nu,\quad n_\mu=(1,\hat{e}_r).
\label{dSKS}
\end{equation}
The gauge field obtained via the Kerr-Schild single copy is
\begin{equation}
A_\mu=\rho\, r^2 n_\mu,
\label{AmudS}
\end{equation}
where we have replaced $\lambda\rightarrow \rho$. We can interpret the
latter parameter by noting that the electrostatic potential in the
Kerr-Schild gauge satisfies
\begin{equation}
\nabla^2A_0=\frac{1}{r^2}\frac{\partial}{\partial r}\left(r^2
\frac{\partial A_0}{\partial r}\right)=6\rho.
\label{phidS}
\end{equation}
Thus, $\rho$ plays the role of a uniform charge density, filling all
space. This is exactly what one expects from the single copy of the
cosmological constant, given that the latter is a uniform energy
density. If one chooses the fiducial metric to be Minkowski rather
than de Sitter space, the conformal coupling in the biadjoint scalar
theory would be absent (due to the vanishing Ricci scalar), but one must
then include the uniform charge density explicitly.

\section{Discussion}
\label{sec:conclude}

In this paper, we have extended our investigation of the double copy
for classical solutions, which commenced in
ref.~\cite{Monteiro:2014cda}. In particular, we studied the Taub-NUT
metric, which goes beyond the results of ref.~\cite{Monteiro:2014cda}
in having a double Kerr-Schild form. The single copy of this solution
a dyon whose electric and magnetic charge copy to mass and NUT charge
respectively in the gravity theory. A similar story exists in higher
spacetime dimensions, for the generalised Plebanski metrics of
refs.~\cite{Chen:2006xh,Chen:2007fs}. \\

We also examined the Taub-NUT solution in (Anti)-de Sitter space, and
found that the single copy also works. This is a highly interesting
result, given that this is the first example of a double copy
involving a non-Minkowski background. The zeroth copy also works,
provided one adds a conformal mass term to the theory, which vanishes
in the Minkowski case. This interpretation of the curved space double
copy is somewhat tentative, and more investigation is necessary
(e.g. involving other background geometries, or scattering amplitudes
on curved space).\\

Work on extending the classical double copy further is in progress.

\section*{Acknowledgments}

We thank Henry Tye for suggestions leading to this project, and
J. J. Carrasco, Einan Gardi and David Miller for useful
discussions. CDW is supported by the UK Science and Technology
Facilities Council (STFC) under grant ST/L000446/1, and is perennially
grateful to the Higgs Centre for Theoretical Physics for
hospitality. DOC is supported in part by the STFC consolidated grant
``Particle Physics at the Higgs Centre'', by the National Science
Foundation under grant NSF PHY11-25915, and by the Marie Curie FP7
grant 631370. AL is supported by Conacyt and SEP-DGRI studentships. RM
is supported by a Marie Curie Fellowship and by a JRF at Linacre
College.

\bibliography{refs.bib}
\end{document}